\documentclass[12pt,twoside, draftclsnofoot, onecolumn, journal]{IEEEtran}
\IEEEoverridecommandlockouts
\usepackage{amsfonts}
\usepackage{amsmath}
\usepackage{theorem}
\usepackage{graphicx}
\usepackage{epsfig}
\usepackage{cite}
\usepackage{epstopdf}
\usepackage{algorithm}
\usepackage{algorithmic}
\usepackage{multicol}
\usepackage{amssymb}

\newtheorem{example}{Example}
\newtheorem{theorem}{Theorem}
\newtheorem{lemma}{Lemma}
\newtheorem{proposition}{Proposition}
\newtheorem{corollary}{Corollary}
\newtheorem{definition}{Definition}

\makeatother
\makeatletter

\begin{document}

\title{Contractive Interference Functions and Rates of Convergence of Distributed Power Control Laws}

\author{Hamid Reza Feyzmahdavian, Mikael Johansson, and Themistoklis Charalambous
\thanks{H. R. Feyzmahdavian, M. Johansson, and T. Charalambous are with ACCESS Linnaeus Center, School of Electrical
Engineering, KTH-Royal Institute of Technology, SE-100 44 Stockholm, Sweden.
E-mails: {\{hamidrez, mikaelj, themisc\}@kth.se.}}
}
\maketitle
\markboth{IEEE Transactions on Wireless Communications}%
{Submitted paper}


\begin{abstract}

The standard interference functions introduced by Yates have been very influential on the analysis and design of distributed power control laws. While powerful and versatile, the framework has some drawbacks: the existence of fixed-points has to be established separately, and no guarantees are given on the rate of convergence of the iterates. This paper introduces contractive interference functions, a slight reformulation of the standard interference functions that guarantees the existence and uniqueness of fixed-points along with linear convergence of iterates. We show that many power control laws from the literature are contractive and derive, sometimes for the first time, analytical convergence rate estimates for these algorithms. We also prove that contractive interference functions converge when executed totally asynchronously and, under the assumption that the communication delay is bounded, derive an explicit bound on the convergence time penalty due to increased delay.  Finally, we demonstrate that although standard interference functions are, in general, not contractive, they are all para-contractions with respect to a certain metric. Similar results for two-sided scalable interference functions are also derived.
\end{abstract}

\IEEEpeerreviewmaketitle


\section{Introduction}

Distributed power control (DPC) algorithms such as~\cite{FoM:93} and \cite{Zan:92} have had an enormous influence on modern wireless systems.
The basic algorithm for adjusting transmit powers to meet predefined Signal-to-Interference-and-Noise-Ratio (SINR) targets can be written as a linear iteration and has been thoroughly analyzed using tools from linear algebra and positive linear systems. In particular, when the SINR targets are feasible, the algorithm converges to a unique fixed-point at a linear rate, i.e., the distance between the iterates and the optimal power allocation decays exponentially. These results can be derived using Perron-Frobenius theory for positive matrices or, alternatively, by showing that the linear iteration is a contraction mapping in a weighted maximum norm (see for example~\cite{Mitra:93}).

An elegant axiomatic framework for studying more general power control iterations was proposed by Yates~\cite{Yat:95}. The so-called \emph{standard interference functions} include the linear iterations, and several important nonlinear power control laws.  Various extensions of the basic framework have been proposed in the literature with the most prominent those by Sung and Leung~\cite{SuL:05} and Schubert and Boche~\cite{Boche:10}. While several results exist for synchronous and asynchronous convergence of interference function iterations, very few results on the convergence rates of such algorithms have appeared in the literature (see for example~\cite{HuY:98,HeC:00,ChT:06} for exceptions). The current proofs are tailor-made and the link to contraction mappings, that has been so powerful in the analysis of the linear iterations, is disturbingly absent. This paper tries to fill this gap.

Contrary to claims in the literature, we demonstrate that interference functions are, in general, not necessarily contraction mappings. However, we show that a slight modification of the scalability axiom of standard interference function allows to guarantee contractivity of the iterations and hence unique fixed-points and linear convergence rates. This condition is satisfied by the basic DPC algorithm and allows to recover the same convergence rate that comes out of a tailored analysis. It also allows to estimate the convergence rate of the other power control schemes considered by Yates, as well as the utility-based power control scheme developed in~\cite{XSH:03}. We also demonstrate how a logarithmic change-of-variables render interference functions para-contractions.  Interestingly, this is the same change of variables that has been used extensively in resource allocation for interference-limited systems (\emph{e.g.},~\cite{JXB:03, Chi:05}). Furthermore, we show that our conditions are also satisfied for the two-sided scalable interference functions introduced in~\cite{SuL:05}, and introduce conditions that guarantee that two-sided scalable interference functions define contraction mappings, and hence have unique fixed-points and linear convergence rates.  Finally, we discuss how asynchronous convergence can be established in our framework.

\subsection{Notation}
Throughout the paper, vectors are written in bold lower case letters and matrices in upper case letters. The set of non-negative real numbers is denoted by $\mathbb{R}_+$. $x_i$ denotes the $i^{th}$ component of a vector $\bold x$, and the notation $\bold x\geq \bold 0$ means that all of the components of $\bold x$ are greater than or equal to zero. The inequality $\bold x\geq \bold y$ implies that $x_i\geq y_i$ for all components $i$. We use $\bold e^{\bold x}$ and $\bold \ln(\bold x)$ to denote component-wise exponential and logarithm of the entries $\bold x$, respectively. The matrix $A$ is said to be \textit{Hurwitz} if all its eigenvalues have negative real parts. It is \textit{Metzler} if all its off-diagonal entries are
non-negative. Given a vector $\bold v>\bold 0$, $\|\cdot\|_\infty^{\text{v}}$ stands for the weighted maximum norm, i.e., $\|\bold x\|_\infty^{\text{v}}=\max_i\left|{{x_i}\over{v_i}}\right|$. The vector norm $\|\cdot\|_\infty^{\text{v}}$ induces a matrix norm, also denoted by $\|\cdot\|_\infty^{\text{v}}$, defined by
\begin{align*}
\| A \|_\infty^{\text{v}} = \max_{\mathbf{x}\neq 0}\frac{\| A\mathbf{x} \|_\infty^{\text{v}}}{\| \mathbf{x} \|_\infty^{\text{v}}}\;.
\end{align*}
When $v_i=1$ for all $i$, we suppress the superscript $\textup{v}$. The spectral radius of a matrix $A$ is the largest magnitude of the eigenvalues of $A$ and is denoted by $\rho(A)$. A sequence $\{\bold x_{n}\}\in \mathbb{R}^{K}$ is said to converge linearly to $\bold x^{\star}$ if there exists a constant $c\in(0,1)$ such that
\begin{align*}
\lim_{n\rightarrow \infty}\frac{\|x(n+1)-x^{\star}\|}{\|x(n)-x^{\star}\|}=c,
\end{align*}
where $\|\cdot\|$ is some norm on $\mathbb{R}^{K}$. If $c=1$, then the sequence is said to converge sub-linearly.

\section{Fixed-point theory and interference functions}

\subsection{Fixed-points, contractions and para-contractions}

We consider iterative algorithms on the form
\begin{align}
\bold x(n+1)&=\bold T\bigl(\bold x(n)\bigr),\;\;\;n=0, 1, 2, \ldots,
\label{Iter}
\end{align}
where $T$ is a mapping from a subset $X$ of $\mathbb{R}^{K}$ into itself. A vector $\bold x^{\star}$ is called a fixed point of
$T$ if $\bold T(\bold x^{\star})=\bold x^{\star}$. If $T$ is continuous at $\bold x^{\star}$ and the sequence $\{\bold x(n)\}$ converges to $\bold x^{\star}$, then $\bold x^{\star}$
is a fixed point of $T$~{\cite[Chapter\;3]{BeT:89}}. Therefore, the iteration \eqref{Iter} can be viewed as an algorithm for finding such a fixed point. $T$ is called a \textit{contraction mapping}, if it has the following property
\begin{equation*}
\|\bold T(\bold x)-\bold T(\bold y)\|\leq c\;\|\bold x-\bold y\|\;,\forall x,y\in X,
\end{equation*}
where $\|\cdot\|$ is some norm on $X$, and $c\in[0,1)$. The following proposition shows that
contraction mappings have unique fixed points and linear convergence rates.
\begin{proposition}[Convergence of Contracting Iterations {\cite[Chapter\;3]{BeT:89}}] If $T:X \rightarrow X$ is a contraction mapping and that $X$ is a closed subset of $\mathbb{R}^{K}$, then:
\begin{itemize}
\item 	(Existence and Uniqueness of Fixed Points) The mapping $T$ has a unique fixed point $\bold x^{\star} \in X$.
\item 	(Linear Convergence) For every initial vector $\bold x(0)\in X$, the sequence $\{\bold x(n)\}$ generated by $\bold x(n+1)=\bold T\bigl(\bold x(n)\bigr)$ converges to $\bold x^{\star}$ linearly. In particular,
\begin {equation*}
\|\bold x(n)-\bold x^{\star}\|\leq c^{n}\|\bold x(0)-\bold x^{\star}\|\;.
\end {equation*}
\end{itemize}
\label{contraction}
\end{proposition}
An operator $T$ on $X$ is called \emph{para-contraction} if
\begin{align*}
\|\bold T(\bold x)-\bold T(\bold y)\|<\|\bold x-\bold y\|\;,\;\;\mbox{for all}\;\; \bold x\neq \bold y\;.
\end{align*}
Para-contractions have at most one fixed point and, in contrast to contractions, may not have a fixed point. As an example, consider the para-contracting function $T(x)=x+e^{-x}$ in $[0, \infty)$. It is easily seen that $T$ has no fixed point. The following theorem summarizes properties of para-contractions.
\begin{proposition}[\cite{edelstein:62}] If $T:X \rightarrow X$ is a para-contraction, then:
\begin{itemize}
\item 	If $T$ has a fixed point $\bold x^{\star}$, then that fixed point is unique; moreover
\item 	If $X$ is a finite-dimensional space, for every initial vector $\bold x(0)\in X$, the sequence $\{\bold x(n)\}$ generated by $\bold x(n+1)=\bold T\bigl(\bold x(n)\bigr)$ converges to $\bold x^{\star}$ .
\end{itemize}
\label{Para}
\end{proposition}
As can be seen from Proposition \ref{Para}, para-contractivity does not yield any estimate of the rate of convergence to the fixed point.


\subsection{Standard interference functions}

The standard interference functions were introduced by Yates~\cite{Yat:95} to study various extensions of the basic distributed power control problem.

\begin{definition}[Standard Interference Function~\cite{Yat:95}]
A function $I:\mathbb{R}^{K}_+ \rightarrow \mathbb{R}^{K}_+$ is called a \emph{standard interference function},  if for all $\bold p\geq \bold 0$ the following properties are satisfied:
\begin{itemize}
\item \textit{Positivity}: $\bold I(\bold p)> \bold 0$\;.
\item \textit{Monotonicity}: If $\bold p\geq \bold p'$, then $\bold I(\bold p)\geq \bold I(\bold p')$\;.
\item \textit{Scalability}: For all $\alpha > 1$, $\alpha \bold I(\bold p) > \bold I(\alpha \bold p)$.
\end{itemize}
\end{definition}

The main convergence result for standard interference functions can be summarized as follows:

\begin{proposition}[\cite{Yat:95}]\label{prop:standard_convergence}
Let $I$ be a standard interference function and consider the iteration
\begin{align}
	\mathbf{p}(n+1) &= \mathbf{I}\bigl(\mathbf{p}(n)\bigr). \label{eqn:interference_iteration}
\end{align}
 Then, if \eqref{eqn:interference_iteration} has a fixed-point, this fixed-point is unique and the iterates $\{ \mathbf{p}(n)\}$ produced by \eqref{eqn:interference_iteration} converge to the fixed-point from any initial vector $\mathbf{p}(0)$.
\end{proposition}

Note that contrary to the result for contraction mappings, the existence of fixed-points has to be verified separately, and no guarantees about the convergence rate of the iterates to the fixed-point are given. Already this should raise the suspicion that standard interference functions do not define contraction mappings. The following simple example establishes that this suspicion is indeed correct.
\begin{example}[Standard interference functions are not contractive] \label{ex:standard_divergent}
Consider $\mathbf{I}(\mathbf{p})= 2\mathbf{p}+\bold 1$ where $\bold 1$ is the vector with all components equal to $1$.
This is a standard interference function, but $\Vert \mathbf{I}(\mathbf{p})-\mathbf{I}(\mathbf{p}^{\prime})\Vert = 2 \Vert \mathbf{p}-\mathbf{p}^{\prime}\Vert$, so it is neither contractive nor para-contractive.
\end{example}

The interference function in Example~\ref{ex:standard_divergent} does not have a fixed-point in the positive orthant, hence does not contradict Proposition~\ref{prop:standard_convergence}. The next example shows that even a standard interference function has a positive fixed point, the iteration~\eqref{eqn:interference_iteration} may converge at a sub-linear rate.

\begin{example}\label{example2} Consider $I(p)=\frac{4}{1+e^{-(p-2)}}$ in $[0,\infty)$. $I(p)$ is a standard interference function since it satisfies positivity, scalability, and monotonicity. $I(2)=2$, so $p^{\star}=2$ is a fixed point. It is easy to verify that $I'(2)=1$, $I''(2)=0$, and $I^{(3)}(2)=-\frac{1}{2}$. Therefore, the iteration~\eqref{eqn:interference_iteration} converges sub-linearly to $p^{\star}$ \cite[Theorem 4]{Stewart:95}.
\end{example}
Example \ref{example2} provides motivations for seeking stronger conditions than standard interference function to ensure contractivity, hence linear convergence rates of the iterations.
%
To this end, one could certainly make a separate analysis of contractivity of the particular interference functions at hand. However, if one can prove contractivity, particularly in the weighted maximum norm, then the interference function framework brings little additional value. The beauty of the framework lies in the easily verifiable conditions that guarantee synchronous and asynchronous convergence.  Next, we will show that a slight reformulation of the scalability condition ensures contractivity.

\section{Contractive interference functions}

We propose to study a class of interference functions which we call \emph{contractive}.

\begin{definition}
A function $I:\mathbb{R}^{K}_+ \rightarrow \mathbb{R}^{K}_+$ is said to be a \textit{contractive interference function} if it, for all $\bold p\geq \bold 0$, satisfies the following conditions:
\begin{itemize}
\item \textit{Positivity}: $\bold I(\bold p)>\bold 0\;.$
\item \textit{Monotonicity}: If $\bold p\geq \bold p'$, then $\bold I(\bold p)\geq \bold I(\bold p')\;.$
\item \textit{Contractivity}: There exists a constant $c\in[0,1)$ and a vector $\bold{v}>\bold 0$ such that for all $\epsilon > 0$, $\bold I(\bold p+\epsilon\bold v) \leq \bold I(\bold p)+c\epsilon\bold v$.
\end{itemize}
\end{definition}
Note that the two first conditions are the same as for standard interference functions, but the scalability condition has now been replaced by contractivity. The following example shows that contractivity, in general, does not imply scalability.
\begin{example}
Consider the function
\begin{align*}
	{I}(p) &=\left\{\begin{array}{ll}
p^2+\frac{1}{100}, & 0\leq p \leq \frac{1}{4},\\
\frac{1}{2}p-\frac{1}{16}+\frac{1}{100}, &\frac{1}{4}\leq p,
\end{array}  \right.
\end{align*}
which is contractive $(c=\frac{1}{2})$. However, the scalability property does not hold for $\alpha=2$ and $p=\frac{1}{8}$ since $I(\alpha p)\not<\alpha I(p)$.
\end{example}
However, with the additional requirement that $I_i : \mathbb{R}^{K}_{+} \rightarrow \mathbb{R}_+$ is concave for all $i=1,\ldots,K$, the contractive interference function is also standard. This result follows immediately from the fact that positivity and concavity imply scalability \cite{Boche:08}.

As shown in the next theorem, contractive interference functions define contraction mappings, which implies that the associated iterations \eqref{eqn:interference_iteration} have unique fixed-points and linear convergence rates.
\begin{theorem}\label{thm:contractive}
If $I$ is a contractive interference function, then it has a unique fixed point $\mathbf{p}^{\star}$ and for every initial vector $\mathbf{p}(0)$, the sequence $\bold p(n+1)=\bold I\bigl(\bold p(n)\bigr)$ converges linearly to $\bold p^{\star}$ as follows
$\|\bold p(n)-\bold p^{\star}\|_{\infty}^{\textup{v}} \leq c^{n}\|\bold p(0)-\bold p^{\star}\|_{\infty}^{\textup{v}}$.
\label{contractivity}
\end{theorem}
\begin{IEEEproof}
Let $\bold p\neq\bold p'$. Since $|p_i-p'_i|\leq \|\bold p-\bold p'\|_{\infty}^{\textup{v}}v_i$ for all $i$, we have
\begin{eqnarray*}
\bold p&=&\bold p'+\bold p-\bold p'\\
&\leq&\bold p'+\|\bold p-\bold p'\|_\infty^{\text{v}}\;\bold{v}\;.
\end{eqnarray*}
Note that $\|\bold p-\bold p'\|_\infty^{\text{v}}>0$. The monotonicity and contractivity properties imply
\begin{eqnarray*}
\bold I(\bold p)&\leq& \bold I(\bold p'+\|\bold p-\bold p'\|_\infty^{\text{v}}\;\bold{v})\\
&\leq&\bold I(\bold p')+c\|\bold p-\bold p'\|_\infty^{\text{v}}\;\bold{v}\;.
\end{eqnarray*}
By interchanging the roles of $\bold p$ and $\bold p'$, $\bold I(\bold p')\leq\bold I(\bold p)+c\|\bold p-\bold p'\|_\infty^{\text{v}}\;\bold{v}$. So for all components of $\bold I(\bold p)$, we have
$|I_i(\bold p)-I_i(\bold p')|\leq c\|\bold p-\bold p'\|_\infty^{\text{v}} \text{v}_i$ which implies that $\|\bold I(\bold p')-\bold I(\bold p)\|_\infty^{\text{v}}\leq c\|\bold p-\bold p'\|_\infty^{\text{v}}$. Therefore, according to Proposition \ref{contraction}, $I$ is contractive and the result follows.
\end{IEEEproof}

To emphasize the modulus $c$ of the contraction mapping, we say that a function is a \emph{c-contractive interference function}. Note that convergence rate is directly related to the number of iterations required for the algorithm to converge. Specifically, if we define the convergence time of the iteration $T_{\delta}$ as the the smallest $t$ such that $\Vert \bold p(t)-\bold p^{\star}\Vert_{\infty}^{\text{v}}\leq \delta$, then, with $R_0=\Vert \bold p(0)-\bold p^{\star}\Vert_{\infty}^{\text{v}}$, $\Vert \bold p(t)-\bold p^{\star}\Vert_{\infty}^{\text{v}} \leq c^{t}R_0$. So if $c<1$, $T_{\delta} \leq\frac{1}{\ln c}\ln\frac{\delta}{R_0}$. We can see that the convergence time goes to $\infty$ as $c$ tends to one.

To show that the concept of contractive interference functions is useful, we will now show that it is readily applied to several distributed power control algorithms from the literature. We assume $K$ users and $R$ base stations and a common radio channel.

First, consider \textit{fixed assignment} interference functions
\begin{align}
I_i(\bold p)&=\gamma_i{\sum_{j\neq i}G_{r_ij}p_j+\eta_{r_i} \over G_{r_i i}},\;\;\;i=1, \ldots, K\;,\label{eqn:linear_interference_function}
\end{align}
where $r_i$ is the user $i$'s base station, $G_{rj}$ is the link gain between base $r$ and user $j$, $\gamma_i$ is the target SINR of user $i$, and $\eta_r$ is the background noise at base $r$. Under fixed assignment, each user $i$ has an assigned base station $r_i\in\{1,\ldots,R\}$. Equation~\eqref{eqn:linear_interference_function} can be rewritten as
\begin{align*}
I_i(\bold p)={\sum_{j=1}^{K}M^{(r_i)}_{ij}p_j+N^{(r_i)}_i},\;\;\;i=1, \ldots, K\;,
\end{align*}
where $N^{(r_i)}_i=\gamma_i \frac{\eta_{r_i}}{G_{r_i i}}$, and
\begin{align}
M^{(r_i)}_{ij}=\left\{\begin{array}{ll}
{\gamma_iG_{r_i j}\over G_{r_i i}}, & j\neq i, \\
\hspace{0.2cm}0, & j=i.
\end{array}  \right.\label{eqn:M}
\end{align}
Define $M$ as an $K\times K$ matrix that has $M^{(r_i)}_{ij}$ as its elements. We then have the following result.
\begin{theorem}
If $\Vert M\Vert_{\infty}^{\textup{v}}<1$ for some $\bold{v}>\bold{0}$, then linear interference functions \eqref{eqn:linear_interference_function} are c-contractive interference functions with $c=\Vert M\Vert_{\infty}^{\textup{v}}$.\label{Prop : Linear}
\end{theorem}
\begin{IEEEproof}
It is clear that linear interference functions are positive and monotone. Furthermore,
\begin{eqnarray*}
I_i(\bold p+\epsilon\bold v)&=&{\sum_{j=1}^{K}M^{(r_i)}_{ij}p_j+N^{(r_i)}_i}+\epsilon\sum_{j=1}^{K}M^{(r_i)}_{ij}v_j\\
&\leq&I_i(\bold p)+\epsilon\Vert M\Vert_{\infty}^{\textup{v}}v_i\;,
\end{eqnarray*}
where we used Proposition \ref{weightednorm}$(\mbox{b})$ in the appendix to get the last inequality. Hence, they are also contractive with $c=\Vert M\Vert_{\infty}^{\textup{v}}$.
\end{IEEEproof}

Since matrix $M$ is a nonnegative square matrix, the following Proposition holds.

\begin{proposition}\textbf{(\cite[Ch. 2]{BeT:89})}\label{weightednorm}
Let $A\in \mathbb{R}^{n \times n}$ be a nonnegative matrix. Then
\begin{itemize}
\item[(a)] There exists a positive vector $\bold v$ such that $\Vert A\Vert_{\infty}^{\textup{v}}<~1$ if and only if $\rho(A)<1$.
\item[(b)] Let $\bold v$ be a positive vector. Then $\sum_{j=1}^n A_{ij}v_j\leq \Vert A\Vert_{\infty}^{\textup{v}} v_i$ for all $i$.
\end{itemize}
\end{proposition}

According to Proposition \ref{weightednorm}, $\rho(M)<1$ is a necessary and sufficient condition for the existence of a positive vector $\bold v$ for which $\|M\|_\infty^{\text{v}}<1$. Any vector of the form $(I-M)^{-1}\bold x$ where $\bold x>\bold 0$ can then be chosen to be $\bold v$~\cite[Proposition~1]{Rantzer:11}. If  the matrix $M$ is irreducible, which is often a reasonable assumption (since we are not considering totally isolated groups of links that do not interact with each other),  it is worth noticing that the positive right Perron-Frobenius eigenvector $\bold v$ is such that $\rho(M)=\Vert M\Vert_{\infty}^{\textup{v}}$ (\cite[Proposition 6.6]{BeT:89}). In either case, Theorem~\ref{Prop : Linear}  confirms that if $\rho(M)<1$, then the fixed assignment iteration has a unique fixed point and a linear convergence rate.  That is, there exists a set of powers to which all transmitters converge exponentially fast, such that all transmitters meet their QoS requirements (minimum SINR for successful reception). This also coincides with previous tailor-made analyses (for example \cite{FoM:93, Mitra:93, Ber:03}).

The convergence rate that can be guaranteed using Theorem~\ref{thm:contractive} depends on the choice of $\mathbf{v}$, and hence on the norm in which we require the iterations to be contractive. In many cases, sufficient but more easily verifiable conditions can be derived by considering $\mathbf{v}=\mathbf{1}$.  For the linear interference function iterations, $\mathbf{v}=\mathbf{1}$ yields the condition that $\Vert M\Vert _{\infty}<1$, i.e.,
\begin{align*}
\frac{G_{r_ii}}{\sum_{j\neq i}{G_{r_i j}}}>\gamma_{i}
\quad \forall \ i = 1,2, \ldots K.
\end{align*}
Since $\rho(M)\leq\Vert M\Vert _{\infty}$, this condition is more conservative, but it is easily verifiable as we only need to check if each of the row sums of matrix $M$ are less than $1$.

We now verify that the minimum power assignment and macro-diversity interference functions from~\cite{Yat:95} are also contractive under appropriate assumptions.
\begin{itemize}
\item {\textit{Minimum Power Assignment}}: At each step of this iterative algorithm, user $i$ is assigned to the base station $r$ at which its transmitted power is minimized. In this case we have
\begin{align*}
I_i(\bold p)=\min_{r\in \{1,\ldots,R\}}\left\{\sum_{j=1}^{K}M^{(r)}_{ij}p_j+N_i^{(r)}\right\},
\end{align*}
where $M^{(r)}_{ij}$ and $N_i^{(r)}$ are defined as \eqref{eqn:M}. Define matrices $M_1,\ldots,M_R$ where the $(i,j)^{th}$ element of $M_r$ is equal to $M^{(r)}_{ij}$. $M_r$ is the normalized gain matrix when all users are assigned to base station $r$. Let $\bold v> \bold 0$ and $c=~\max_r\{\Vert M_r\Vert_{\infty}^{\textup{v}}\}$. Then
\begin{eqnarray*}
I_i(\bold p+\epsilon\bold v)&=&\min_r\left\{{\sum_{j=1}^{K}M^{(r)}_{ij}p_j+N_i^{(r)}+\sum_{j=1}^{K}M^{(r)}_{ij}v_j}\right\}\\
&\leq&\min_r\left\{\sum_{j=1}^{K}M_{ij}^{(r)}p_j+N_i^{(r)}+\epsilon \Vert M_r\Vert_{\infty}^{\textup{v}} v_i\right\}\\
&\leq &I_i(\bold p)+c\epsilon v_i\;.
\end{eqnarray*}
At this point we have shown that if a common positive vector $\bold v$ exists such that $\Vert M_r\Vert_{\infty}^{\textup{v}}<1$ for all $r$, then the interference function is contractive. The question now would be how to check for the existence of such a $\bold v>\bold 0$, and this will be addressed now. Let ${L}$ be a set containing all possible mappings $l:\{1,\ldots,K\}\rightarrow \{1,\ldots,R\}$. Each mapping $l$ represents the allocation of users to base stations and $l(i)$ denotes that under mapping $l$, user $i$ is assigned to base station $l(i)$. We can construct the normalized gain matrix under assignment $l$, $M^l$, by letting the $i^{th}$ row of $M^l$ equalt the $i^{th}$ row of $M_{l(i)}$. The following lemma provides a test for the existence of a common $\bold v>\bold 0$.
\begin{lemma}\label{lemma2}
Let $M^l$ be the normalized gain matrix under assignment $l\in L$. If $\rho(M^l)<1$ for all~$l$, then there exists a positive vector $\bold v$ such that $\Vert M_r\Vert_{\infty}^{\textup{v}}<~1$ for all $r$.
\end{lemma}
\begin{IEEEproof}
Since $M^l$ is a nonnegative matrix with zero diagonal entries, then if  $\rho(M^l)<1$ for all $l$, all eigenvalues of $M^l$ are strictly inside the unit circle. Thus, the real part of $\lambda_i(M^l-I)$ is negative for all $i$, and hence $M^l-I$ is Hurwitz. On the other hand, the off-diagonal elements of $M^l-I$ are nonnegative; therefore, $M^l-I$ is also Metzler. As a result, there exists a positive vector $\bold v$ such that $(M_r-~I)\bold v <0$ for all $r$~\cite[Theorem 4]{Knorn:09}. Therefore, $M_r\bold v < \bold v$ holds, which implies that $\Vert M_r\Vert_{\infty}^{\textup{v}}<~1$ for all $r$.
\end{IEEEproof}

By Lemma \ref{lemma2}, contractivity of minimum power assignment iteration boils down to computing the spectral radius of $R^K$ matrices $M^l$. A more conservative condition can be derived by considering $\bold v=\bold 1$. Then, $I$ is contractive if $\Vert M_r\Vert_{\infty}<1$ for all $r$. In this case, one is only required to calculate the row sums of $R$ matrices to establish contractivity.

\item {\textit{Macro-Diversity}}: The interference function is defined as
\begin{align*}
I_i(\bold p)={1 \over {\sum_{r=1}^{R} {1 \over {\sum_{j=1}^{K}M^{(r)}_{ij}p_j+N^{(r)}_i}}}}\;.
\end{align*}
Similar to minimum power assignment iteration, we can define $R$ matrices $M_1\ldots,M_R$. Let $\bold v> \bold 0$ and $c=\max_r\{\Vert M_r\Vert_{\infty}^{\textup{v}}\}$. We have
\small
\begin{eqnarray*}
I_i(\bold p+\epsilon\bold v)-I_i(\bold p)&=&{{{\sum_{r=1}^{R} {{\sum_{j=1}^{K}\epsilon M^{(r)}_{ij}v_j} \over \left({\sum_{j=1}^{K}M^{(r)}_{ij}p_j+N^{(r)}_i}\right) \left({\sum_{j=1}^{K}M^{(r)}_{ij}(p_j+\epsilon v_j)+N^{(r)}_i}\right)}}}\over {{\sum_{r=1}^{R} {1 \over {\sum_{j=1}^{K}M^{(r)}_{ij}p_j+N^{(r)}_i}}\times\sum_{r=1}^{R}{1 \over {\sum_{j=1}^{K}M^{(r)}_{ij}(p_j+\epsilon v_j)+N^{(r)}_i}}}}}\\
&\leq&  {{{\sum_{r=1}^{R} {{\epsilon \Vert M_r\Vert_{\infty}^{\textup{v}}v_i} \over \left({\sum_{j=1}^{K}M^{(r)}_{ij}p_j+N^{(r)}_i}\right)\left({\sum_{j=1}^{K}M^{(r)}_{ij}(p_j+\epsilon v_j)+N^{(r)}_i}\right)}}}\over {{\sum_{r=1}^{R} {1 \over {\sum_{j=1}^{K}M^{(r)}_{ij}p_j+N^{(r)}_i}}\times\sum_{r=1}^{R}{1 \over {\sum_{j=1}^{K}M^{(r)}_{ij}(p_j+\epsilon v_j)+N^{(r)}_i}}}}}\\
&\leq&c\epsilon v_i\;.
\end{eqnarray*}
\normalsize
Thus, if $c<1$ for some $\bold v>\bold 0$, then the interference function is contractive. By Lemma \ref{lemma2}, $\rho(M^l)<1$ for all $l \in L$ guarantees contractivity.

While the convergence rate estimates for the interference functions that we have considered earlier are tight or appear to be tight, the situation for the macro diversity results is less clear.
Assuming that a user's power contributes to its own interference, Hanly~\cite{Hanly:96} showed that the macro-diversity interference function has a unique fixed point
if $\sum_{i=1}^K\gamma_i<R$. This condition is insensitive to the channel gains and hence
to the position of base stations. While we are unaware of any convergence rate estimates for the original or modified macro diversity interference functions, a related interference function has been investigated by Rodriguez \emph{et. al}~\cite{Rodriguez:08}. To understand the relationship with our results, let  $m_{r,i}(\bold p)={\sum_{j\neq i}G_{rj}p_j}$ and re-write the macro-diversity interference function as
\begin{align*}
I_i(\bold p)={\gamma_i \over {\sum_{r=1}^R {{G_{ri}} \over m_{r,i}(\bold p)+\eta_r}}}
.\end{align*}
By introducing $\widehat{m}_{i}(\bold p)=\max_r\{m_{r,i}(\bold p)\}$ and $\widehat{\eta}=\max_r\{\eta_r\}$,
\begin{align*}
\overline{I}_i(\bold p)=\gamma_i{{\widehat{m}_{i}(\bold p)+\widehat{\eta}} \over {\sum_{r=1}^R {{G_{ri}}}}}
\end{align*}
is a strict overestimate of the original macro diversity interference function. For this interference function, some results have been obtained in~\cite{Rodriguez:08}. It is readily verified that $\overline{I}_i(\bold p)$ is contractive, since for any $\epsilon>0$, we have
\begin{eqnarray*}
\overline{I}_i(\bold p+\epsilon\bold v)-\overline{I}_i(\bold p)&=&
\gamma_i{ \max_r\left\{{\sum_{j\neq i}G_{rj}(p_j+\epsilon v_j)}\right\} - \max_r\left\{{\sum_{j\neq i}G_{rj}p_j}\right\} \over {\sum_{r=1}^R {{G_{ri}} }}}\nonumber\\
&\leq&\epsilon\gamma_i{ \max_r\left\{{\sum_{j\neq i}G_{rj}v_j}\right\} \over {\sum_{r=1}^R {{G_{ri}} }}}.
\end{eqnarray*}
This inequality can be rewritten as
\begin{eqnarray}
\overline{I}_i(\bold p+\epsilon\bold v)-\overline{I}_i(\bold p)
&\leq&\epsilon \max_r\left\{{\sum_{j=1}^K H_{ij}^{(r)}v_j}\right\}\;,\label{macro1}
\end{eqnarray}
where
\begin{align}
H^{(r)}_{ij}=\left\{\begin{array}{ll}
{\gamma_i G_{rj}\over \sum_{r=1}^R G_{ri}}, & j\neq i, \\
\hspace{0.2cm}0, & j=i.
\end{array}  \right.\nonumber
\end{align}

Define matrix $H_r$ with $(i,j)^{th}$ element equal to $H_{ij}^{(r)}$. Let $c=\max_r\{\Vert H_r\Vert_{\infty}^{\textup{v}}\}$, then Inequality~\eqref{macro1} becomes $\overline{I}_i(\bold p+\epsilon\bold v)-\overline{I}_i(\bold p)\leq c \epsilon v_i$. Hence, if $\rho(H^{(l)})<1$ for all $l \in L$, then by Lemma \ref{lemma2}, $\Vert H_r\Vert_{\infty}^{\textup{v}}<1$ for all $r$ and the interference function is contractive. The result derived in~\cite{Rodriguez:08} is $\gamma_i {\sum_{j\neq i} G_{ri} \over {\sum_{r=1}^R {{G_{ri}} }}}<1$ for all $i,r$, which is equivalent to $\|H_r\|_\infty<1$ for all $r$. This indicates that the stability condition in~\cite{Rodriguez:08} is more conservative than our result.

\end{itemize}

To show that our framework allows to go beyond the known results, consider the utility-based power control (UBPC) from~\cite{XSH:03}. The associated interference function is
\begin{align}
I_i^u(\bold p)&=\Biggl({\sum_{j\neq i}G_{rj}p_j+\eta_r \over G_{ri}}\Biggr)f_i^{-1}\Biggl(\alpha_i{\sum_{j\neq i}G_{rj}p_j+\eta_r \over G_{ri}}\Biggr), \label{eqn:ubpc1}
\end{align}
where $\alpha_i$ is a price coefficient and $f^{-1}_i(x)$ is a decreasing function on $[{\underline{K}}_i, {\overline{K}}_i]$ for all $i$ given by $f_i(SIR_i)=U'_i(SIR_i)$ where $U_i$ is a utility function of user $i$. In their paper, Xiao~\emph{et al.} use a sigmoidal utility function
\begin{align}
U_i(SIR_i)&={1 \over {1+e^{-a_i(SIR_i-b_i)}}}, \label{eqn:ubpc3}
\end{align}
where $b_i=\gamma_i-a_{i}^{-1}\ln(a_i\gamma_i-1)$. Let $M^{(r)}_{ij}$ be defined as \eqref{eqn:M} and let $M_b$ be a matrix with $(i,j)^{th}$ element equal to $b_iM^{(r)}_{ij}$. We will next show that the framework of contractive interference functions will allow us to analytically bound the convergence rate, which has an immediate use for tuning the algorithm parameters. Specifically, we have the following result.
\begin{theorem}\label{UBPC}
Consider the interference function $I^{u}$ defined in~\eqref{eqn:ubpc1}--\eqref{eqn:ubpc3}. If $c~= \rho(M_b) <1$, then $I^{u}$ is a c-contractive interference function.
\end{theorem}
\begin{IEEEproof}
Let $\bold v > \bold 0$. For all $\epsilon>0$ we have
\small
\begin{align}
I_i^u(\bold p+\epsilon\bold v)&=\Biggl({\sum_{j=1}^{K}M^{(r)}_{ij}(p_j+\epsilon v_j)+N^{(r)}_i}\Biggr)
f_i^{-1}\left(\alpha_i\biggl({\sum_{j=1}^{K}M^{(r)}_{ij}(p_j+\epsilon v_j)+N^{(r)}_i}\biggr)\right)\nonumber\\
&<\Biggl({\sum_{j=1}^{K}M^{(r)}_{ij}(p_j+\epsilon v_j)+N^{(r)}_i}\Biggr)
f_i^{-1}\left(\alpha_i\biggl({\sum_{j=1}^{K}M^{(r)}_{ij}p_j+N^{(r)}_i}\biggr)\right)\nonumber\\
&=\Biggl({\sum_{j=1}^{K}M^{(r)}_{ij}p_j+N^{(r)}_i}\Biggr)f_i^{-1}\left(\alpha_i\biggl({\sum_{j=1}^{K}M^{(r)}_{ij}p_j+N^{(r)}_i}\biggr)\right)\nonumber\\
&\hspace{5mm}+\epsilon\Biggl({\sum_{j=1}^{K}M^{(r)}_{ij}v_j}\Biggr)f_i^{-1}\left(\alpha_i\biggl({\sum_{j=1}^{K}M^{(r)}_{ij}p_j+N^{(r)}_i}\biggr)\right)\nonumber\\
&=I_i^u(\bold p)+\epsilon\Biggl({\sum_{j=1}^{K}M^{(r)}_{ij}v_j}\Biggr)f_i^{-1}\left(\alpha_i\biggl({\sum_{j=1}^{K}M^{(r)}_{ij}p_j+N^{(r)}_i}\biggr)\right),\label{UBPC1}
\end{align}
\normalsize
where the first inequality comes from the fact that $f^{-1}_i(x)$ is a decreasing function. Under the sigmoidal utility function \eqref{eqn:ubpc3}, the maximum value of $U'_i(SIR_i)$ occurs at point $b_i$, so $\max f^{-1}_i(x)=\arg\max_x f_i(x)=b_i$. Thus, Inequality \eqref{UBPC1} becomes
\begin{align*}
I_i^u(\bold p+\epsilon\bold v) &\leq I_i^u(\bold p)+\epsilon \Biggl({\sum_{j=1}^{K}b_iM_{ij}v_j}\Biggr)\nonumber\\
&\leq I_i^u(\bold p)+\epsilon\Vert M_b\Vert_{\infty}^{\textup{v}}\;.
\end{align*}
Hence, if $\Vert M_b\Vert_{\infty}^{\textup{v}}<1$, then UBPC iteration linearly converges to a unique fixed point and the result follows.
\end{IEEEproof}

The following numerical example illustrates our result.
\begin{example}
We consider UBPC under the simulation scenario described in~\cite{XSH:03}. Here, four mobiles share a channel with link gain matrix $G$ given by
\small
\begin{align*}
G=\begin{bmatrix} 10^{-4} & 6.82\times 10^{-7} & 3.57\times 10^{-8} & 2.12\times 10^{-8}\\
1.52\times 10^{-7} & 6.25\times 10^{-4} & 3.51\times 10^{-6} & 1.98\times 10^{-7}\\
7.67\times 10^{-9} & 2.44\times 10^{-8} & 1.23\times 10^{-6} & 5.16\times 10^{-9}\\
2.63\times 10^{-7} & 4.82\times 10^{-8} & 2.56\times 10^{-7} & 3.28\times 10^{-5}
\end{bmatrix}\;.
\end{align*}
\normalsize
The noise power is $0.5$ and the target SIRs of the users are $6$, $6$, $8$ and $10$ dB, respectively. We assume that four users in the
system use sigmoidal utility function with parameters $1.02$, $1.32$, $0.88$ and $1.05$, respectively. The price coefficient $\alpha$ is equal to $5000$ for all users.
Fig.~\ref{fig:UBPC_simulation} shows a comparison of the norm between power vector and optimal power vector versus number of iterations for UBPC algorithm and the theoretical bound obtained from Theorem \ref{UBPC}. It can easily be seen that the distance between the iterates of UBPC algorithm and optimal power allocation at iteration $n$, $\|\bold p(n)-\bold p^{\star}\|_{\infty}^{\textup{v}}$, is always less than $c^{n}\|\bold p(0)-\bold p^{\star}\|_{\infty}^{\textup{v}}$ and decays exponentially.

\begin{figure}
\centering
\includegraphics [height=2.0 in, width=2.5 in]{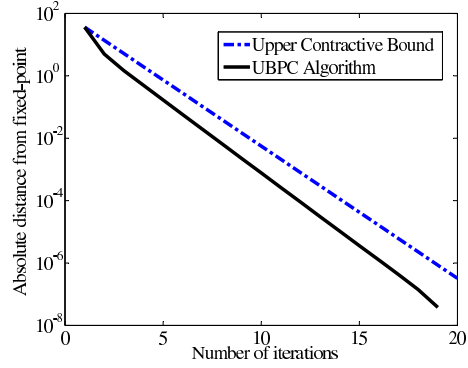}
\caption{Comparison of upper bound on convergence rate of UBPC obtained in Theorem \ref{UBPC} and the actual convergence rate of UBPC for the scenario considered in example 4.} \label{fig:UBPC_simulation}
\end{figure}

\end{example}


Another useful result shows that imposing an upper and lower bound on a contractive interference function does not change the contractivity properties.
\begin{theorem}
If $I$ is a c-contractive interference function, then so is
\begin{align*}
\mathbf{I}^q(\mathbf{p}) = \max\bigl\{ {\mathbf p}_{\rm min}, \min\{ {\mathbf p}_{\rm max}, \mathbf{I}(\bold p)\}\bigr\}.
\end{align*}
\end{theorem}
\begin{IEEEproof}
$\mathbf{I}^q(\bold p)$ satisfies positivity and monotonicity. It remains to show that $\bold I^q(\bold p)$ satisfies contractivity. Let $\epsilon>0$. The contractivity condition of $\bold I(\bold p)$ implies
\begin{eqnarray}
\mathbf{I}^q(\bold p+\epsilon\bold v)&=&\max\bigl\{\bold p_{\min}, \min\{\bold p_{\max}, \bold I(\bold p+\epsilon\bold v)\}\bigr\}\nonumber\\
&\leq&\max\bigl\{\bold p_{\min}, \min\{\bold p_{\max}, \bold I(\bold p)+c\epsilon\bold v\}\bigr\}\nonumber\\
&\leq&\max\bigl\{\bold p_{\min}, \min\{\bold p_{\max}, \bold I(\bold p)\}+c\epsilon\bold v\bigr\}\nonumber\\
&\leq&\max\bigl\{\bold p_{\min}, \min\{\bold p_{\max}, \bold I(\bold p)\}\bigr\}+c\epsilon\bold v\nonumber\;.
\end{eqnarray}
Therefore, $\mathbf{I}^q(\bold p+\epsilon\bold v)\leq \mathbf{I}^q(\bold p)+c\epsilon\bold v$.
\end{IEEEproof}

Finally, we consider the distributed robust power control algorithm (DRPC) from~\cite{Yang:11} . Denote the normalized channel gain between user $i$'s base station and user $j$ as $M_{ij}^{(r)}=\overline{M}_{ij}^{(r)}+\triangle M_{ij}^{(r)}$, where $\overline{M}_{ij}^{(r)}$ is the nominal value, and $\triangle M_{ij}^{(r)}$ is the perturbation associated with $\overline{M}_{ij}^{(r)}$. The uncertainty set of $\mathcal{M}$ is
\begin{align*}
\mathcal{M}=\{M\;|\;M_i\in\mathcal{M}_i,\;i=1,\ldots,K\},
\end{align*}
where $M_i$ is the $i^{th}$ row of matrix $M=[M_{ij}^{(r)}]$, and $\mathcal{M}_i$ is the uncertainty set of $M_i$.
The interference function of DRPC is defined by
\begin{eqnarray*}
I_i(\bold p)=\sup_{M_i \in \mathcal{M}_i} \left(M_i^T \bold p\right)+N_i,\;\;i=1, \ldots, K.
\end{eqnarray*}
The convergence rate of DRPC algorithm is verified by the following theorem.
\begin{theorem}\label{DRPC}
If $c=\sup_{M \in \mathcal{M}} \Vert M\Vert_{\infty}^{\textup{v}}<1$ for some $\bold{v}>\bold{0}$, then DRPC interference function is c-contractive.
\end{theorem}
\begin{IEEEproof}
Similar to Theorem \ref{Prop : Linear}.
\end{IEEEproof}

Note that if $\mathcal{M}$ is bounded and $\sup_{M \in \mathcal{M}}\rho(M)<1$, then there exists a $\bold v>\bold 0$ for which $\sup_{M \in \mathcal{M}} M\bold v< \bold v$~\cite[Theorem 1]{Yang:11}. Thus, Theorem \ref{DRPC} coincides with the stability condition derived in~\cite{Yang:11}.

\section{Interference functions and para-contractions}
We have already shown that standard interference functions do not define contraction mappings. However, the convergence results for standard interference functions are identical to those of para-contractions in Proposition \ref{Para}, so there should be a link between the two. This section shows one such link. In particular, we will demonstrate that a logarithmic change of variables $\mathbf{s}=\ln(\mathbf{p})$ makes the iterations para-contracting in the new variables. Interestingly, this is the same change-of-variables that has been very useful in convexifying various resource allocation problems for interference-limited wireless systems ~\cite{JXB:03, Chi:05}.

\begin{theorem}
Suppose that the interference function $I$ is standard. Then the change of variables $\bold s= \bold \ln(\bold p)$ and $\widetilde{\bold I}({\bold s})=\bold \ln\bigl(\bold I(\bold e^{ {\bold s}})\bigr)$ transforms the interference function $I:\bold p \rightarrow \bold I(\bold p)$ into a para-contracting function $\widetilde{I}:\bold s \rightarrow\widetilde{\bold I}({\bold s})$.
\label{paracontraction}
\end{theorem}

\begin{IEEEproof} We rewrite the properties of standard interference function in the new coordinates.
\begin{enumerate}
\item[\textbf{A1}] If $\bold s_1\leq\bold s_2$, then $\bold e^{ \bold s_1}\leq\bold e^{ \bold s_2}$. By monotonicity property of $\bold I(\bold p)$
\begin{eqnarray*}
\bold I(\bold e^{ \bold s_1})\leq \bold I(\bold e^{ \bold s_2})&\Rightarrow&\bold \ln\bigl(\bold I(\bold e^{ \bold s_1})\bigr)\leq \ln\bigl(\bold I(\bold e^{ \bold s_2})\bigr)\\
&\Rightarrow&\widetilde{\bold I}(\bold s_1)\leq \widetilde{\bold I}(\bold s_2).
\end{eqnarray*}
\item[\textbf{A2}] For any $\epsilon>0$ and by using scalability property of $\bold I(\bold p)$, we have
\begin{eqnarray*}
\widetilde{\bold I}(\bold s+ \epsilon \textbf{1})&=&\ln\bigl(\bold I(e^{\epsilon}\bold e^{ \bold s})\bigr)\\
&<&\ln\bigl(e^{\epsilon} \bold I\left(\bold e^{ \bold s}\right)\bigr)\\
&=&\epsilon\textbf{1}+\widetilde{\bold I}(\bold s).
\end{eqnarray*}
\end{enumerate}
For any $\bold s_1,\bold s_2\in\mathbb{R}^{n}$, we have
\begin{eqnarray*}
\bold s_1&\leq&\bold s_2+\|\bold s_1-\bold s_2\|_\infty\textbf{1}.
\end{eqnarray*}
Let $\bold s_1\neq \bold s_2$. By properties A1-A2 of $\widetilde{\bold I}(\bold s)$
\begin{eqnarray*}
\widetilde{\bold I}(\bold s_1)&\leq& \widetilde{\bold I}(\bold s_2+\|\bold s_1-\bold s_2\|_\infty\textbf{1})\\
&<&\widetilde{\bold I}(\bold s_2)+\|\bold s_1-\bold s_2\|_\infty\textbf{1}.
\end{eqnarray*}
By interchanging the roles of $\bold s_1$ and $\bold s_2$ in the preceding inequality, we obtain
\begin{align*}
\widetilde{\bold I}(\bold s_2)<\widetilde{\bold I}(\bold s_1)+\|\bold s_1-\bold s_2\|_\infty\textbf{1}.
\end{align*}
Consequently for all components of $\widetilde{\bold I}(\bold s)$, we can write
\begin{equation*}
|\widetilde{I}_i(\bold s_1)-\widetilde{I}_i(\bold s_2)|< \|\bold s_1-\bold s_2\|_\infty\;, \; i=1,\ldots,K,
\end{equation*}
which implies that $\|\widetilde{\bold I}(\bold s_1)-\widetilde{\bold I}(\bold s_2)\|_\infty<\|\bold s_1-\bold s_2\|_\infty$. Therefore, $\widetilde{\bold I}({\bold s})$ is para-contracting.
\end{IEEEproof}

Theorem \ref{paracontraction} helps us to understand that interference functions are para-contractions with respect to a certain metric. Specifically, we note the following:
\begin{corollary}
Standard interference functions are para-contractions with respect to the metric
\begin{align*}
	d_c(\mathbf{p},\mathbf{p}^{\prime}) &= \max_i \left\vert\ln \frac{p_i}{p_i^{\prime}}\right\vert\;.
\end{align*}
\end{corollary}
It is important to note that standard interference functions are para-contractions on the metric space induced by $d_c$, irrespective of whether they have a fixed-point or not. To guarantee convergence of the iterates, we must verify that the iteration has fixed-points. The following theorem can then be useful.
\begin{theorem}
Given a standard interference function $I$, if there exists a $\bold p'$ such that $\bold I(\bold p')\leq\bold p'$, then a fixed point exists.
\end{theorem}
\begin{IEEEproof}
Let $X=\{\bold p: \bold 0\leq\bold p\leq \bold p'\}$. By the positivity and monotonicity properties, we have $\bold 0< \bold I(\bold p) \leq \bold p'$. Hence, $I$ maps $X$ to itself. Let us
define function $f: X\mapsto [0,\infty]$ by $f(\bold p)=d_c\bigl(\bold p, \bold I(\bold p)\bigr)$. Since $f$ is continuous and $X$ is compact, $f$ attains its minimum. Let $f(\widetilde{\bold p})=\min_{\bold p\in X}f(\bold p)$. If $\widetilde{\bold p}$ is not a fixed point of  $I$, then by para-contractivity
\begin{eqnarray*}
f\bigl(\bold I(\widetilde{\bold p})\bigr)&=&d_c\biggl(\bold I(\widetilde{\bold p}), \bold I\bigl(\bold I(\widetilde{\bold p})\bigr)\biggr)\\
&<& d_c\bigl(\widetilde{\bold p}, \bold I\bigl(\widetilde{\bold p})\bigr)=f(\widetilde{\bold p}).
\end{eqnarray*}
contradicting the fact that $\widetilde{\bold p}$ minimizes $f$. Hence $\bold I(\widetilde{\bold p})=\widetilde{\bold p}$.
\end{IEEEproof}


\section{Two-sided interference functions}

Sung and Leung~\cite{SuL:05} present a new class of functions called \emph{two-sided scalable interference functions} which generalizes the standard interference functions to allow for simple and powerful analysis of certain opportunistic power control laws:
\begin{definition}[Two-sided scalable interference functions~\cite{SuL:05}]
A function$I:\mathbb{R}^{K}_+ \rightarrow \mathbb{R}^{K}_+$ is called a \emph{two-sided scalable interference function},  if for all $\bold p\geq \bold 0$, $\bold I(\bold p)$ satisfies:
\begin{itemize}
\item \textit{Positivity}: $\bold I(\bold p)> \bold 0$.
\item \textit{Two-sided scalability}: For all $\alpha>1$, $\frac{1}{\alpha}\bold p\leq\bold p'\leq\alpha\bold p$ implies
$\frac{1}{\alpha}\bold I(\bold p)<\bold I(\bold p')<\alpha\bold I(\bold p)$.
\end{itemize}
\end{definition}
Note how monotonicity and scalability conditions of standard interference function have been replaced by the two-sided scalability condition. However, every standard interference function is also two-sided scalable. The key convergence result reads as follows:
\begin{proposition}[\cite{SuL:05}]
Let $I$ be a two-sided scalable interference function and consider the iteration \eqref{eqn:interference_iteration}. If the iteration has a fixed-point $\bold p^{\star}$, then this fixed-point is unique and the sequence $\{\mathbf{p}(n)\}$ generated by the iteration converges to $\bold p^{\star}$ for every initial value $\mathbf{p}(0)$.
\end{proposition}
The convergence conditions for two-sided scalable interference functions coincide with those of para-contractions. In particular, neither existence of fixed-points nor any convergence rates are guaranteed. To guarantee existence and uniqueness of fixed-points along with convergence rates, the natural concept is to consider \emph{two-sided contractive interference functions}.
\begin{definition}
A function $I:\mathbb{R}^{K}_+ \rightarrow \mathbb{R}^{K}_+$ is called \emph{two-sided contractive interference function} if it, for all $\mathbf{p}\geq \bold 0$ satisfies
\begin{itemize}
\item Positivity: $\mathbf{I}(\mathbf{p})>\bold 0$.
\item Two-sided contractivity: There exists a constant $c\in[0,1)$ and a vector $\bold{v}>\bold 0$ such that for all $\epsilon>0$, $\bold p'-\epsilon\bold v\leq\bold p\leq\bold p'+\epsilon\bold v$ implies that $\bold I(\bold p')-c\epsilon\bold v\leq\bold I(\bold p)\leq\bold I(\bold p')+c\epsilon\bold v$.
\end{itemize}
\end{definition}

The associated convergence theorem now reads
\begin{theorem}
If $I$ is a two-sided contractive interference function, then \eqref{eqn:interference_iteration} has a unique fixed point $\mathbf{p}^{\star}$ and the sequence $\{ \mathbf{p}(n)\}$ generated by the iteration \eqref{eqn:interference_iteration} converges linearly to $\mathbf{p}^{\star}$ from every initial value $\mathbf{p}(0)$.
\end{theorem}
The proof follows similarly to the contractive interference function proof and is omitted in this paper.
Next, we show how two-sided scalable functions relate to para-contractions.
\begin{corollary}
Suppose that $I:\bold p \rightarrow \bold I(\bold p)$ is two-sided scalable. Then $I$ is a para-contraction with respect to the metric $d_c(\bold p, \bold p')=\max_i\vert \ln{p_i \over p'_i} \vert$ .
\end{corollary}

Our results in this section are related to the work by M{\"o}ller and J{\"o}nsson~\cite{Moller:2009},  who studied stability of higher-order power control laws. They demonstrated that in logarithmic variables, two-sided scalability implies global Lipschitz continuity of the interference function, and an alternative restriction allows to establish convergence rates and uniqueness of fixed-points.


\section{Asynchronous Power Control}
So far, we have examined synchronous power control algorithms. In this case, every component of the vector $\bold p$ is updated at every time step, using information of the transmit powers used by all transmitters in the previous iteration. However, a nice feature of the standard interference functions is that they also guarantee convergence in the absence of synchronization. In this section, we will demonstrate that contractive interference functions also converge asynchronously.

Asynchronous computation models may be divided into \emph{totally asynchronous} and \emph{partially asynchronous} ~\cite[Chapter 6--7]{BeT:89}. Let $T$ be the set of times when some transmitter updates its power, and let $T^i\subseteq T$ be the times when transmitter $i$ executes an update. To model that the transmitter might need to update its power using old information from other transmitters, let $\tau_{j}^{i}(t)$ be the time at which the most recent version of $p_j$ available to node $i$ at time $t$ was computed. Node $i$ executes the update
\begin{align}
p_i(t+1) =\begin{cases}
       I_i\bigl (p_1({\tau}_{1}^i(t)),\cdots,p_n({\tau}_{n}^i(t))\bigr ), & \forall t\in T^i,\\
       p_i(t), & \forall t\not\in T^i.
	\end{cases} \label{eqn:asynchronous_iteration}
\end{align}

\begin{definition}(\textit{Total Asynchronism} ~\cite[Chapter 6]{BeT:89})
The iteration \eqref{eqn:asynchronous_iteration} is \emph{totally asynchronous} if the sets $T^i$ are infinite for all $i$, and if $\{t_k\}$ is a sequence of elements of $T^i$ that tends to infinity, then it also holds that $\lim_{k\to \infty}{\tau}_{j}^i(t_k)=\infty$ for all $j$.
\end{definition}
Loosely speaking, this assumption guarantees that no transmitter ceases to update its power, and that such updates eventually propagate to all other transmitters in the network. Yates showed that if an iteration involving standard interference functions converges synchronously, it also converges when it is executed totally asynchronously. A similar result holds for contracting interference functions:

\begin{theorem}
Let $I$ be a contractive interference function. Then, the iterates produced by \eqref{eqn:asynchronous_iteration} converge to the unique fixed-point under total asynchronism.
\end{theorem}

This result is proven by noticing that contractive interference functions define max-norm contractions which converge under total asynchronism (\emph{e.g.},~\cite[Page 434]{BeT:89}). A similar result can be established for two-sided contractive interference functions using the same arguments.

While the totally asynchronous convergence result is comforting, it does not quantify the longer convergence times one could expect with increasing information delays. An additional advantage of the contractive interference function framework is that such results can be developed.
Note that in the totally asynchronous model, the delays can become unbounded as $t$ increases. Consider, in contrast, a situation where all mobiles update their powers at each iteration but the information delay is guaranteed to be bounded. In particular, there exists a positive integer $D$ such that $t-D<\tau_j^i(t)\leq t$ for all $i$ and $j$.  The following theorem gives a bound on the convergence rate of contractive interference functions under this model of asynchronicity.
\begin{theorem}
If $I:\mathbb{R}^{K} \rightarrow \mathbb{R}^{K}$ is c-contractive, then from any initial power vector $\mathbf{p}(0)$, the asynchronous power control algorithm satisfies:
\begin{align*}
\|\bold p(n)-\bold p^{\star}\|^{\textup{v}}_{\infty} \leq \overline{c}\;^n \|\bold p(0)-\bold p^{\star}\|^{\textup{v}}_{\infty},
\end{align*}
where $\overline{c}=c^{{1 \over {D+1}}}$.
\end{theorem}
\begin{IEEEproof}
Contractivity of $I$ implies that $\frac{|I_i(p)-p^{\star}_i|}{v_i} \leq c \max_{j\in\{1, \cdots, K\}} \frac{|p_j-p_j^{\star}|}{v_j},\;i=1,\ldots,K$. Let $d_{i}$ be a maximum communication delay between user $i$ and
other users, then the convergence rate of the asynchronous algorithm \eqref{eqn:asynchronous_iteration} is the unique solution of the equation ~\cite[pp.\;442]{BeT:89}
\begin{align}
\rho=\max \{c\rho^{-d_1},\cdots , c\rho^{-d_K}\}.\label{eqn:rate asynchronous_iteration}
\end{align}
Since $D=\max\{d_1, \cdots, d_K\}$, Equation~\eqref{eqn:rate asynchronous_iteration} can be written as $\rho=c\rho^{-D}$ and by letting $\overline{c}=\rho$, the proof is complete.
\end{IEEEproof}
Note that contractive interference functions converge under arbitrary bounded delays, and that our result provides an explicit bound on the impact that an increasing delay has on the convergence rate. Moreover, computing the convergence time for the asynchronous update, we find that
\begin{align*}
	T_{\delta} &\leq (D+1)\frac{1}{\ln c}\ln \frac{\delta}{R_0}.
\end{align*}
Hence, a communication delay of $D$ results in a convergence time that is no more than $D+1$ longer than that of the ideal synchronous iteration.

Wu \emph{et al}~\cite[Theorem 2]{Yang:11}, using the definition for the spectral radius, derive the convergence rate of the robust interference function they consider, when updated every $K$ steps (corresponding to $D+1$ steps in our setup). More specifically, the convergence rate $E$ for the case with no uncertainties, is given by
\begin{align}
E\geq \frac{1}{K}\log\frac{1}{\rho(M)},
\end{align}
which is equivalent to our derived convergence rate. When, however, there exists reduced message passing and increased robustness consideration, this is incorporated in matrix $M$, thus reducing the rate of convergence. Therefore, their derived bounds on the convergence rates for the cases they consider, justify our analysis and constitute a worked example of our proposed framework.

\section{Conclusions}

This paper has explored the connections between the standard interference function framework and the theory for fixed-point iterations. We have shown that standard interference functions do not define contraction mappings and introduced contractive interference functions, a slight modification of the standard interference functions, that guarantee existence and uniqueness of fixed-point along with linear convergence of iterates. We have demonstrated that several important distributed power control algorithms proposed in the literature are contractive and derived the associated convergence rates. In some cases, such as linear iterations, the convergence rate coincides with known results from the literature that has been obtained using a detailed and tailored analysis. In other cases, such as the utility-based power control, we provide the first convergence rate estimates in the literature. This paper also provided a link between standard interference functions and para-contractions. This link involves a logarithmic change of variables (alternatively, analysing the iterations with respect to a specific metric), which coincides with the change-of-variables that has been so successful in convexifying resource allocation problems in interference-limited wireless systems. Associated results for two-sided scalable interference functions have also been given.

\bibliography{interference}

\end{document}